\documentclass{article}

\input{tcilatex}
\begin{document}

\begin{center}
\bigskip {\Large A comment on "A dimensionless measure for adhesion and
effects of the range of adhesion in contacts of nominally flat surfaces" by
M. H. Muser}

\bigskip

M.Ciavarella

Politecnico di BARI. DMMM dept. V Orabona, 4, 70126 Bari.

Hamburg University of Technology, Dep Mech Eng, Am Schwarzenberg-Campus 1,
21073 Hamburg, Germany

email: mciava@poliba.it
\end{center}

\bigskip Muser attempts in a recent paper (Muser, 2016) to introduce single
dimensionless parameters to razionalize the complex problem of the contact
of rough multiscale surfaces in the presence of adhesion, and checks his
parameters against some numerical simulations with a quite large number of
grid points for today's computational capabilities ($\sim 10^{6}$).

1) The first generalization relates to so called "Tabor parameter". The
original Tabor parameter (Tabor, 1977) which controls the validity of the
DMT ("long-range" adhesion) vs JKR regime ("short-range" adhesion) solution\
for the sphere, is%
\begin{equation}
\mu _{sphere}=\left( \frac{R\Delta \gamma ^{2}}{E^{\ast 2}\Delta r^{3}}%
\right) ^{1/3}  \label{Tabor}
\end{equation}%
where $R$ is the sphere radius, $\Delta \gamma $ is work of adhesion, $%
\Delta r$ is the range of attraction of adhesive forces, and $E^{\ast }$ the
plane strain elastic modulus. Similarly to Persson and Scaraggi (2014),
Muser merely generalizes the Tabor parameter, by considering the relevant
"local" mean radius $R_{c}\simeq 2/h_{rms}^{\prime \prime }$ which largely
depends on the truncating wavevector of roughness $q_{1}$, and denominates
this Tabor parameter $\mu _{T}$. But while the original Tabor parameter is
unambigously defined for the sphere, for the case of multiscale surfaces, we
need to return to the "real" meaning of it, to understand why various
possibilities exist for its generalization, and the dimensional analysis
Muser suggests is just one of many possible alternatives. In essense, the
Tabor parameter defines the limit of validity of the Linear Elastic Fracture
Mechanics, like in the `small-scale yielding' criterion. {Since the elastic
tensile stresses near the boundary of the contact region are given by $%
\sigma (x)=\frac{K_{\mathrm{I}}}{\sqrt{2\pi x}}\;$, where $x$ is the
distance from the boundary, the width of the region where the stress is
higher than the theoretical strength $\sigma _{0}$ is }$s_{0}=\frac{E^{%
\raisebox{0.7mm}{$ *$}}\Delta \gamma }{\pi \sigma _{0}^{2}}\;$, and we
prescribe for JKR to be a good approximation that {$s_{0}\ll a_{\min }$,
where $a_{\min }$ is strictly for a rigorous application of dimensional
analysis, the (*) \textit{smallest length scale associated with the geometry
of the problem}. This could be for example (i) the smallest width of the
contact region (and this leads univoquely to the original Tabor parameter in
the sphere problem, given the dependence on }$R${), or of the separation
region, when the contact is almost complete (see }Ciavarella, et al. 2019).
So, in truth, the "generalized Tabor parameter" is load-dependent, and this
\ limits the generalization $\mu _{T}$ of (Muser, 2016) to the small-load
limit, as it\ should be clearly recognized, or anyway when condition {(*) is
satisfied -- which may not be trivial to check. }

2)\ The second attempt of a dimensionless general quantity is the
"dimensionless surface energy", defined as 
\begin{equation}
\Delta \gamma _{rss}=\frac{\Delta \gamma }{E^{\ast }}\frac{\tanh \left( \mu
_{T}\right) }{\left( h_{rms}^{\prime }\right) ^{3}}
\end{equation}%
where $h_{rms}^{\prime }$ is the root mean-square gradient of the surface, $%
\tanh $ is introduced as an empirical fitting between the "correct"
asymptotics in the two limits of small and large Tabor coefficients $\mu
_{T} $. Unfortunately, this second attempts sums the uncertainties of the
generalization of Tabor parameter, with further ones, like assuming we are
in "unsticky range" and that the area increases due to adhesion by a linear
proportion of the load, and probably other, hidden, ones. The positive check
with some numerical results suffers from the fact that the numerical results
of Muser correspond to limited range in the spectrum of roughness (spanning
less than 3 decades, so only reaching from nm to $\mu m$ wavelength size,
while clearly most real surfaces will necessarily have roughness on scales
much larger than this), for which Muser is not to blame since this is
limited by present computational capabilities (similarly narrow spectra are
considered by Pastewka \& Robbins, 2014).

A rigorous physicist would \bigskip immediately try a possible extrapolation
to $\Delta \gamma _{rss}$ when the spectrum becomes infinitely large:
defining "magnification" $\zeta =q_{1}/q_{0}$, where $q_{0}$ is some
reference short wavevector truncation, and $q_{1}$ is the large wavevector
truncation, and assuming a power law PSD spectrum for simplicity as it is
common, $h_{rms}^{\prime }\sim \zeta ^{1-H}$, while $h_{rms}^{\prime \prime
}\sim \zeta ^{2-H},$ where $H$ is Hurst exponent which is $H=3-D$ where $D$
is fractal dimension of the surface) and hence $\tanh \left( \mu _{T}\right)
\sim $ $\left( h_{rms}^{\prime \prime }\right) ^{-1/3}$ so that 
\begin{equation}
\Delta \gamma _{rrs}\sim \zeta ^{\left( 2-H\right) \frac{2}{3}-3+3H}\sim
\zeta ^{\left( 7H-5\right) /3},\qquad \zeta \rightarrow \infty
\end{equation}
which goes to zero if $\left( 7H-5\right) /3<0$ or $H<5/7=\allowbreak 0.714$%
. However, we know from alternative semi-analytical investigations which
permit to explore wide range of PSD\ spectra (Joe et al., 2018), that
surfaces may be sticky or non-sticky even in this limit but independently on 
$H$, as the contact solution converges and doesn't depend on the "local"
quantities $h_{rms}^{\prime \prime },h_{rms}^{\prime }$, that the reference
to "DMT" or "JKR" regimes is misleading since the "generalized Tabor
parameter" goes to zero but the problem may remain not defined by a DMT\
analysis, and the main parameters ruling the problems are the macroscopic
quantities rms heights $h_{rms}$ and the short wavevector truncation $q_{0}:$
hence, this \ "dimensionless surface energy" quantity is \textit{meaningless}%
.

Notice incidentally that similar apparent contradictions occur in the
Pastewka-Robbins's suggested criterion for stickiness based on numerical
observations {(see }Ciavarella, et al. 2019), which can be cast in the form 
\begin{equation}
\zeta ^{(1-5H/3)}<C\;,  \label{PR2}
\end{equation}%
where $C$ is a positive constant, which in the limit $\zeta \rightarrow
\infty $, would seems to suggest similar results than Muser (2016), except
for $H<0.6$ (and the difference in the limit $H$ may be due to different
ways of extrapolating numerical results or to different precisions).
Viceversa, {Violano \textit{et al.} (2019) suggest that for low fractal
dimension $(D\simeq 2.2)$ rough hard surfaces stick for }

\begin{equation}
\frac{h_{rms}}{\Delta r}<\left( \frac{9}{4}\frac{\sigma _{0}/E^{\ast }}{%
\Delta rq_{0}}\right) ^{3/5}  \label{VIOLANO3}
\end{equation}%
which naturally and simply corresponds to well known empirical Dalhquist
criterion (Dalhquist, 1969a, 1969b), which for 50 and more years has simply
postulated that effective adhesives should have elastic modulus lower than
about $1$ MPa, regardless of the local $h_{rms}^{\prime \prime
},h_{rms}^{\prime }$ values of roughness, which would anyway be extremely
difficult to measure in real life. Similar results were obtained in a much
simpler theory (Ciavarella, 2018, Ciavarella \&\ Papangelo, 2019) which
incidentally fit Pastewka-Robbins' (2014) results regardless of the local $%
h_{rms}^{\prime \prime },h_{rms}^{\prime }$ which seem so crucial in the
Muser (2016) (and also Pastewka-Robbins' 2014) numerical interpolations.

\section{\protect\bigskip References}

Barber, J. R. (2018). Contact mechanics (Vol. 250). Springer. Berlin.

Ciavarella, M. (2018). A very simple estimate of adhesion of hard solids
with rough surfaces based on a bearing area model. Meccanica, 53(1-2),
241-250.\bigskip

Ciavarella, M., Joe, J., Papangelo, A., \& Barber, J. R. (2019). The role of
adhesion in contact mechanics. Journal of the Royal Society Interface,
16(151), 20180738.

Ciavarella, M., \& Papangelo, A. (2019). Extensions and comparisons of BAM
(Bearing Area Model) for stickiness of hard multiscale randomly rough
surfaces. Tribology International, 133, 263-270.

Ciavarella, M., Xu, Y., \& Jackson, R. L. (2019). The generalized Tabor
parameter for adhesive rough contacts near complete contact. Journal of the
Mechanics and Physics of Solids, 122, 126-140.

Dahlquist, C. A. in Treatise on Adhesion and Adhesives, R. L. Patrick (ed.),
Dekker, New York, 1969,2, 219.

Dahlquist, C., Tack, in Adhesion Fundamentals and Practice. 1969, Gordon and
Breach: New York. p. 143-151.

Joe, J., Thouless, M. D., \& Barber, J. R. (2018). Effect of roughness on
the adhesive tractions between contacting bodies. Journal of the Mechanics
and Physics of Solids, 118, 365-373.

M\"{u}ser, M. H. (2016). A dimensionless measure for adhesion and effects of
the range of adhesion in contacts of nominally flat surfaces. Tribology
International, 100, 41-47.

Pastewka, L., \& Robbins, M. O. (2014). Contact between rough surfaces and a
criterion for macroscopic adhesion. Proceedings of the National Academy of
Sciences, 111(9), 3298-3303.

Persson, \bigskip\ B. N. J. , Scaraggi, M. (2014) Theory of adhesion: Role
of surface roughness, J. Chem. Phys. 141, 124701.

Tabor, D. (1977) Surface forces and surface interactions. J. Colloid
Interface Sci. 58, 2.

Violano, G., Afferrante, G., \& Ciavarella, M. (2018), On stickiness of
multiscale randomly rough surfaces. arXiv preprint arXiv:1810.10960.

\end{document}